\documentstyle[epsfig]{mn2e}
\bigskip

\title{Comparative study between $N$-body and Fokker-Planck simulations for rotating star clusters: {\rm{I}}. Equal-mass system}

\author{Eunhyeuk Kim$^{1}$, Ilsang Yoon$^{1,2}$, Hyung Mok Lee$^1$, and Rainer Spurzem$^3$
\thanks{e-mail:
ekim@astro.snu.ac.kr}\\
$^1$FPRD, Department of Physics and Astronomy, Seoul National University, Seoul 151-742, Korea\\
$^2$Department of Astronomy, University of Massachusetts, Amherst, MA 01002,
USA\\
$^3$Astronomische Rechen-Institut,  Zentr. Astron. Univ. Heidelberg
(ZAH), Monchhofstrasse 12-14, 69120 Heidelberg, Germany}

\begin{document}
\maketitle

\begin{abstract}
We have carried out $N$-body simulations for rotating star
clusters with equal mass and compared the results
with Fokker-Planck models. These two
different approaches are found to produce fairly similar results,
although there are some differences with regard to the detailed aspects.
We confirmed the acceleration of the core
collapse of a cluster due to an initial non-zero angular momentum
and found a similar evolutionary trend in the central density and
velocity dispersion in both simulations. The degree of acceleration
depends on the initial angular momentum. Angular momentum is
being lost from the cluster due to the evaporation of stars with a large angular
momentum on a relaxation time scale.
\end{abstract}


\section{Introduction}
\medskip

There are two different approaches to study the dynamical evolution
of collisional stellar systems (for example, globular clusters):
statistical approach and direct integration of the $N$-body equations
of motion. Among the statistical methods, the Fokker-Planck (henceforth
referred as FP) equation, which is a second order approximation of the collisional Boltzmann
equation, has been frequently used. The FP equation is solved by using either
Monte Carlo techniques (e.g., the series of eight papers, from
Spitzer \& Hart 1971 to Spitzer \& Mathieu 1980, see also
an alternative approach by H\'enon 1971, and for recent adaptations
Giersz 1998, Joshi, Rasio, \& Portegies Zwart 2000, Freitag \& Benz 2001),
or the direct numerical solution of the discretized FP equation on a mesh.
In this paper we focus on the latter approach. One-dimensional
(1D) FP models
assuming a spherical symmetry and isotropic velocity dispersion
have been extensively exploited
during the last several decades, and they have successfully elucidated
the full dynamical history of star clusters (Cohn 1980;
Drukier et al. 1992). Two-dimensional (2D) anisotropic models are a
generalization of the 1D model, in which the anisotropy of the
velocity dispersion between the radial and the tangential directions is
taken into account; however, the isotropized distribution
function continues to be used for determining the diffusion coefficients
(e.g., Takahashi 1995, 1996, 1997).

The statistical methods have to make several simplifying
approximations and have some limitations. More realistic simulations
can be performed by directly integrating the complete equations of motion of
all stars. However, the $N$-body integration still requires a huge amount of
computing power. An improvement in computing facilities, particularly
the advent of special purpose hardware such as GRAPE machines
(Makino \& Taiji 1998; Makino et al. 2003;
Fukushige, Makino \& Kawai 2005), made it possible to perform a high accuracy $N$-body
simulations with one million body
(Makino \& Funato 2005; Berczik et al. 2006; Iwasawa et al. 2006; Harfst et al. 2006).
The comparisons between the results obtained from 1D or 2D
FP models and direct $N$-body models generally show good agreement (e.g., Spurzem \&
Aarseth 1996). Comparative studies that use the currently available
$N$-body solver for studying the dynamical evolution of collisional stellar
systems are very important for checking the validity and limitations of
the statistical methods. There have been several previous researches involving
comparative studies (Giersz \& Heggie 1994a,b, 1997; Giersz \&
Spurzem 1994; Khalisi, Amaro-Seoane \& Spurzem 2005;
Freitag, Rasio \& Baumgardt 2006) and the comparisons show that for isolated
non-rotating star clusters the results of FP simulations are
generally in good agreement with those of $N$-body simulations. However, when a
tidal boundary is included, a discrepancy between the $N$-body and FP models
arises; this discrepancy becomes sensitive to $N$ because the relaxation and crossing
time scales are related to different dynamical processes
(Takahashi \& Portegies Zwart 1998, 2000; Baumgardt 2001). For
example, the treatment of the tidal boundary has to be performed carefully in
FP models since the mass evaporation process involves both orbital
dynamics and two-body relaxation. The $N$-body model has to imitate
the FP technique to remove stars immediately if they acquire an energy
higher than the tidal energy; with these precautions,
good agreement can be obtained (Spurzem et al. 2005).

Another extension of the 1D FP model was carried out to
include the effects of rotation. The first numerical simulation
of FP models for rotating clusters was pioneered by Goodman (1983),
but neither the results nor the code was published.
A more detailed and extended work by Einsel \& Spurzem (1999, henceforth referred as
Paper I), who developed a new 2D FP code named ``FOPAX'' for this study from scratch,
revealed that rotating clusters collapse faster than
non-rotating ones. A post core collapse study of rotating star clusters,
improving ``FOPAX'' by several features, including
an energy source due to formation and hardening of three-body
binaries,
was done later by Kim et al. (2002, henceforth referred as Paper \rm{II}).
Papers I and II
studied the dynamical evolution of rotating stellar systems  using
the orbit-averaged 2D FP equations only for equal-mass systems.
In both cases and in the study of Goodman (1983), it was assumed that
the distribution function $f$ depends only on the energy and $z$-
component of the angular momentum ($J_z$).
Kim et al. (2004, henceforce referred as Paper
\rm {III}) extended the method to multi-mass systems, and exhibit interesting
results concerning the segregation of mass and angular velocity with heavy stars
in the cluster centres. Fiestas, Kim, \& Spurzem (2006) modelled individual rotating
globular clusters using the model, and Fiestas \& Spurzem (2007) included a
star-accreting central black hole with a loss cone.
In general, according to the strong Jeans theorem there are
three integrals of motion with regard to the orbital motion of stars
in axisymmetric potentials (Binney \& Tremaine 1987). Two of the
integrals of motion are known as $E$ and $J_z$. However, the nature
of the third integral of motion is not known clearly.
In addition to the standard approximation required for FP models,
the 2D FP models presented here and in the Papers \rm{I} to {III} ignore any
effect of the third integral.
For stellar systems not too much flattened, it might be possible to construct 3D FP models
following the three integral models of Lupton \& Gunn (1987), in which the square of
total angular momentum ($J^2$) was adopted as an approximate third integral.
However, it would be an extremely difficult task numerically and physically (owing to the diffusion
coefficients!); even then it would still be approximate since we do not know the
third integral analytically (compare also Binney \& Tremaine (1987) regarding this issue).
Hence, a thorough comparison with direct $N$-body models appears to
be the best method to acquire knowledge on the validity of all approximations made
in our 2D FP models.

There have been some  preliminary studies of rotating star
clusters involving comparisons between $N$-body and FP methods (Boily
2000; Boily \& Spurzem 2000; Ardi, Mineshige \& Spurzem 2006;
Ernst et al. 2007). Although these authors found good
agreements between these two methods, the number of cases studied is
rather limited. This can be understood in two ways. First, for a smaller value of
$N$ (say up to a few $10^4$) a large number of statistically independent
simulations are needed, and only the ensemble average can be compared with the approximate FP
models. On the other hand, different physical models, such as isolated or tidally
limited models, different degree of central concentrations of stars (i.e., different central potential),
need to be studied.
In this paper, we have carried out a series of
numerical $N$-body simulations of rotating stellar systems, which
are directly comparable with our 2D FP models in Papers \rm{I} and
\rm{II}. By comparing the results with those
obtained from FP models, we can investigate the validity of the
assumptions made in rotating FP models.

This paper is organized as follows. In the next section, we briefly
describe initial $N$-body models and compare them with FP models.
In section 3, we present the numerical results obtained from $N$-body
simulations and their comparisons with those of FP models. The summary
and discussions are given in the last section.

\section{The Models}
\medskip
\subsection{Numerical methods}
\medskip

The Fokker-Planck (FP) code FOPAX, which takes into account the effect of rotation,
was developed in Papers I, II, and III in order to study the secular evolution
of star clusters having initial rotation.
%

For performing direct $N$-body simulations to study the dynamical evolution
of rotating stellar systems, we have used the currently
available high accuracy, collisional $N$-body code NBODY6 (Aarseth 1999). The
NBODY6 code uses the fourth-order Hermite scheme with hierarchical block time steps
(HTS) and the Ahmad-Cohen neighbor scheme for particle integration.
Close encounters between stars and persistent binaries formed by three-body
interactions are solved for their internal motion by using two-body regularization
methods (Kustaanheimo \& Stiefel 1965) and chain regularization
(Mikkola \& Aarseth 1990, 1993, 1996, 1998). Although NBODY6 is capable of
dealing with many more astrophysical components such as the existence of
primordial binaries and stellar mass-loss due to stellar
evolution, we have considered only the treatment of close encounters
between stars in the present study since we are mainly interested in
the role of the initial angular momentum on cluster dynamics and
in comparisons with FP simulations.

\subsection{Initial models and boundary condition}
\medskip
In Paper \rm{II}, the initial 2D FP models are generated
according to Lupton \& Gunn (1987). Our initial $N$-body models that
follow rotating King models with a central concentration of $W_0 = 6$
have the same conditions as those of 2D FP models in Paper \rm
{II}. We have constructed the initial models for the $N$-body
simulations from the 2D FP models in Paper \rm {II} by
random number generation. Three different initial rotations
($\omega_0 = 0.0, 0.3$, and $0.6$) are considered in the present
work. In Table 1, we list some information on the initial models used for the present
simulations.


\begin{table}
\begin{minipage}{75mm}
\raggedleft \caption{Initial parameters for the equal-mass models}
\centering \vspace{1truemm}
\begin{tabular}{@{}cccccc@{}} \hline\hline
$W_0$ & $\omega_0$ & $r_t/r_c$ & $r_h/r_t$ & $T_{tot} / |W|$
\footnote{$T_{tot} / |W|$ : ratio of rotational energy to potential energy} & $N$ \\[3pt] \hline\hline
      &     0.0    &  18.0    &  0.15 & 0.000 & 10240\\
6     &     0.3    &  14.5    &  0.18 & 0.035 & 10240\\
      &     0.6    &  9.6     &  0.24 & 0.101 & 10240\\[1pt] \hline
\end{tabular}
\end{minipage}
\label{tab1}
\end{table}

Since most globular clusters are bound to their host galaxy, they are
tidally limited and stars escape from them through a
tidal boundary. There are many previous studies that considered
the effects of the tidal field on cluster evolution and comprehended
it as an important component of the evolution (e.g.,
Lee \& Ostriker 1986; Lee \& Goodman 1995; Takahashi \& Portegies
Zwart 1998; Takahashi \& Lee 2000; Yim \& Lee 2002; Lee et al.
2006; Spurzem et al. 2005). Among the few different implementations of modeling the tidal
effects, we adopted the instantaneous removal of stars whose total
energy exceeded tidal energy of the cluster. This
approximation is known to be inconsistent with the realistic
$N$-body treatment for small $N$ models, but the inconsistency
decreases for a large value of $N$. We considered the equal-mass models,
which are tidally bound to their host galaxy, in order to compare the obtained results with the FP
results in Paper \rm {II}.
We
have modified the original NBODY6 code to imitate the tidal
environment of the clusters modeled with the 2D FP
equation in Paper \rm{II}; this implies that we promptly remove stars
whose energies exceed
the tidal energy ($e > e_{tid}$, see eq. \ref{eq-tid}), as considered
in 2D FP models. In order to maintain the density within
the tidal radius constant, the tidal radius decreases with time
when there is a loss of mass through the boundary. Subsequently, we have adjusted
the tidal energy at every regular time step according to the following expression,

\begin{equation}
e_{tid}\,(t) = - \frac{GM(t)}{r_{tid}} \label{eq-tid}
\end{equation}

where $r_{tid}$ and $M(t)$ are the tidal radius and the total mass of the
cluster within the tidal radius, respectively. We have used the
initial tidal radius obtained from the FP models to compute the mean
density that is kept constant.

The number of stars ($N$) in a cluster is one of the important
parameters for the dynamical evolution of the cluster. While the
computational burden (except for the core-collapse phase) does not significantly depend
on $N$ in statistical FP method, the number of stars is
very important in the $N$-body simulation as the computation time becomes
nearly proportional to $N^3$. We use $N = 10240$ for the present equal-mass
models only because the number should be close to that used in
Paper II ($N=5000$). In testing the validity of our FP models, it does not
matter that the actual number of stars in globular clusters is significantly larger.
The choice of the number of stars determines the relative strength of the
three-body interactions, which initiate the post-collapse phase
(see e.g., Spurzem \& Aarseth 1996).

\begin{figure}
\epsfig{figure=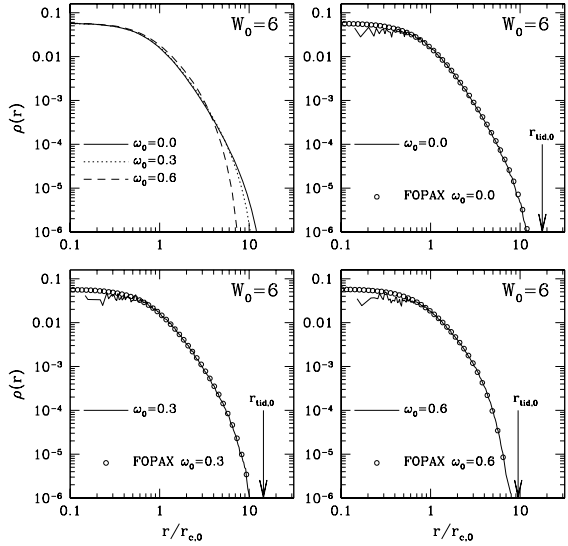, height=0.490\textwidth, width=0.490\textwidth}
\vspace{-3mm}
\caption{Radial profiles of the density for the initial $N$-body models with a central potential of $W_0=6$
and $\omega_0=0.0$, $0.3$, and $0.6$. The upper-left panel shows the FOPAX results.
For comparison, the density profile of the initial FP model with $(W_0,\omega_0)=(6,0.6),(6,0.3),$ and $(6,0.0)$ is shown
(open circles). The $N$-body realization of the initial model shows good agreement with that
of the FP model, except for the central region.}
\label{Fig 1.}
\end{figure}

The realization of the rotating King model
for $N$-body simulations is shown in Figs. 1 and 2. We have shown the
radial profiles of density for both the $N$-body and FP models (Fig. 1) and
the distribution of the radial and tangential velocities of the stars in the $N$-body
realization of the rotating King models (Fig. 2).
Three $N$-body models having different degrees of rotation
are compared with the 2D FP models. Each density profile is
obtained from the mean of 10 different initial models generated by
different random seed numbers. The open circles represent the density
profile adopted in the FP models. Each $r_{tid,0}$ is the initial tidal
radius derived from the FP model. Excellent agreement is observed
between the 
radial profile beyond the core radius ($r_c$) in the $N$-body
realization and that in the FP model. However, within $r_c$, the density of the $N$-body
model is slightly lower than that of FP model. This may be
due to the fact that the number of stars inside the core is rather
small. However, due to inevitable random fluctuations, it is
impossible to construct initial $N$-body models perfectly identical to the FP
models. We believe that statistical FP models agree very well with the averaged
$N$-body models and increasing the number of stars will improve
the degree of agreement. Fig. 1 also shows that increasing the rotation decreases
the concentration of the stellar system (smaller ratio between the
tidal and the core radii).


\begin{figure}
\epsfig{figure=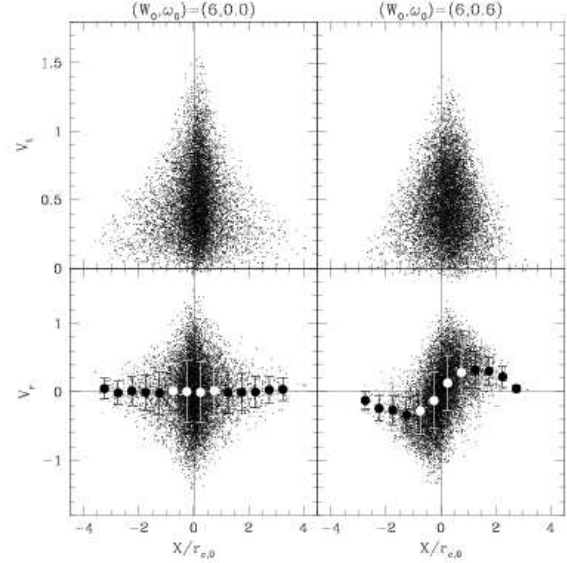, height=0.490\textwidth, width=0.490\textwidth}
\vspace{-3mm}
\caption{Tangential and radial velocity distributions of the initial
$N$-body models with a central potential of $W_0=6$
and initial rotation of $\omega_0=0.0$ (left panels) and $0.6$ (right panels).
The X-axis represents the sky-projected equatorial distance measured in units of
initial core radius. The mean radial velocity
distribution along the projected equator is shown by filled circles with $1\sigma$
errors. The central solid body rotation and the subsequent highly differential rotation
are typical of rotating clusters (see Fig. 10 of Paper \rm{II} for
comparison).}
\label{Fig 2.}
\end{figure}

The position-velocity distributions that are sky-projected are displayed
in Fig. 2 for the $N$-body models of non-rotating ($\omega_0=0.0$, left panels) and
highly rotating ($\omega_0=0.6$, right panels) clusters with the central potential of
$W_0=6$.
To have maximum effect of intial rotation in sky projectied distribution we
project the model clusters on sky in such a way that the rotating axis is a perpendicular axis
in sky-projected plane.
The distance to the rotation axis is measured in units of the initial
core radius. We also show the mean profiles of the radial velocities with
$1\sigma$ errors (filled circles). The profile of the radial velocity for the non-rotating model
shows increasing velocity dispersion toward the center of the cluster.
A typical velocity structure of the rotating model is shown clearly in
the distribution of the radial velocities of the rotating model: rigid-body
rotation inside the core radius and highly differential rotation subsequently.

\section{Results}
\medskip

\subsection{Core collapse, central density, and central velocity dispersion}
\medskip

\begin{figure}
\epsfig{figure=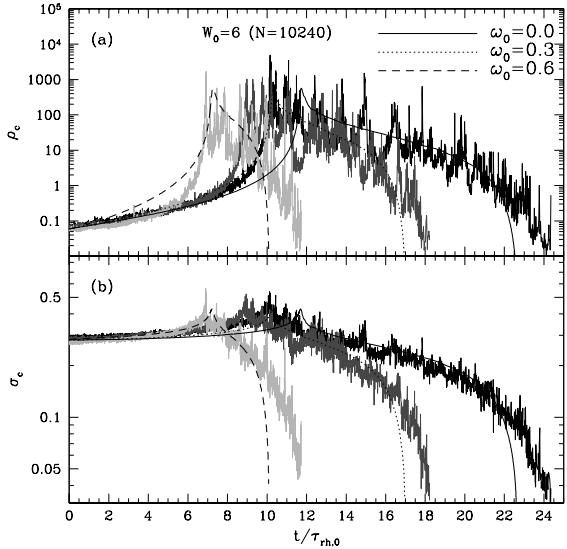, height=0.490\textwidth, width=0.490\textwidth}
\vspace{-3mm}
\caption{Evolution of central density and central velocity dispersion for
each models. FOPAX models are shown with smooth lines.}
\label{Fig 2.}
\end{figure}

\begin{figure}
\epsfig{figure=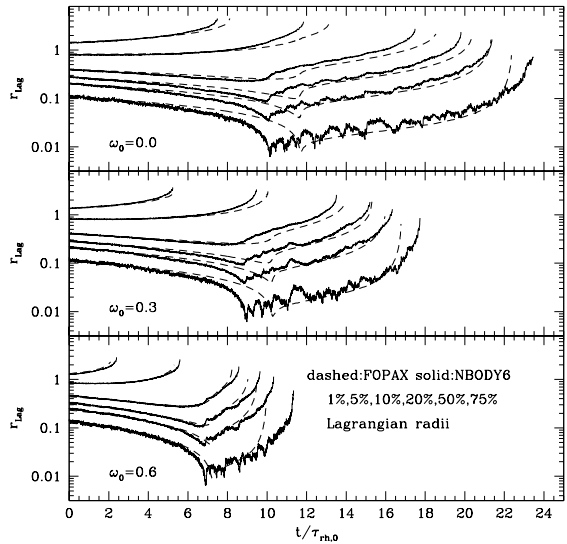, height=0.490\textwidth, width=0.490\textwidth}
\vspace{-3mm}
\caption{Time evolutions of the Lagrangian radii containing $1\%, 5\%, 10\%, 20\%, 50\%$, and
$75\%$ of the initial cluster mass. The initial parameters that
characterize the cluster model are written on the lower-left corner of each panel.
The FOPAX calculations are shown with dashed lines on each panel.}
\label{fig_lagr}
\end{figure}

We start the discussions by presenting the results for the evolution of the
central density and central velocity dispersion (Fig. 3). The central
density increases with time due to the two-body relaxation for an
equal-mass system, where the time is measured in units of initial
half-mass relaxation time ($\tau_{rh,0}$). The expression for
$\tau_{rh,0}$ for the equal-mass system is given by the following
formula (Spitzer \& Hart 1971):

\begin{equation}
\tau_{rh,0} = 0.138  \frac{N^{1/2} r_{h,0}^{3/2}}{G^{1/2} m^{1/2} \rm{ln} \Lambda},
\label{eq-trhi}
\end{equation}
where $N$, $r_{h,0}$, $G$, $m$, and $ \ln \Lambda$ = $\ln (\gamma
N)$ denote the total number of stars, initial half-mass
radius, gravitational constant, mean mass of stars and
Coulomb logarithm, respectively. It has been shown by Giersz \& Spurzem
(1994) and Giersz \& Heggie (1994a,b) that the best agreement
between the direct $N$-body calculations and orbit-averaged FP
equation is achieved when the coefficient $\gamma$ in the Coulomb
logarithm has a value of 0.11. Therefore we use this value in the present work as it is also used in
Paper \rm {II}.
Since the half-mass radius varies with the rotation parameter $\omega_0$,
the values of $\tau_{rh,0}$ could also depend on
$\omega_0$, even for models with the same $W_0$.

The time of core collapse ($t_{cc}$), the time for the complete
disruption of the cluster ($t_{dis}$), and the cluster mass at the
time of core collapse in units of initial cluster mass ($M_{cc}$) are listed
in Table 2. We can easily notice that the rotating models evolve
faster than the non-rotating ones, in both the $N$-body and FP models. The
faster the cluster rotates the shorter the time taken for the core collapse.
As discussed in detail in Papers \rm {I} \& \rm{II}, the acceleration is
caused by the combination of gravothermal and gravogyro instabilities.

In Fig. 3, we show the evolution of the central density ($\rho_c$) and
central velocity dispersion ($\sigma_c$) of the cluster. The results obtained
from the FP and $N$-body models are displayed as smooth lines
and solid lines with a large fluctuation, respectively. Although the core
collapse occurs at slightly different times (one can estimate $t_{cc}$
more accurately in Fig. 4), there is good agreement between the
FP and $N$-body results. The $N$-body model produces rather
noisy data since there is a significant statistical fluctuation in the
physical parameters. However, we can still perform some quantitative
comparisons between the FP and the $N$-body approaches. In the early evolutionary
stage, the central density derived from the $N$-body is less than the value
obtained by FP for the model with $\omega_0=0.6$.
This may reflect the difficulty in determination of the central density for $N$-body
models, especially for rapidly rotating clusters that are significantly flattened.
Therefore, the most rapidly rotating model has the largest discrepancy
with regard to
the central density between the N-body and the FP models.

From Table 2, we notice that the times for core collapse in the FP simulations are
generally greater than those in the N-body simulations by 5--15\%.
Apparently, different approaches should yield some different
results, and we regard a small difference of 5--15\% in $t_{cc}$ as being
insignificant. These differences would decrease
for large $N$ models as the assumptions made in the FP method
become more appropriate. In fact, we have observed that $t_{cc}$ decreases
for model with large $N$ value when we compared the simulations performed for
$N$-body models with $N=5000$ and $N=10240$.

After the core collapse, the evolutions of $\rho_c$ derived from the $N$-body
and FP models are also somewhat different from each other. Toward the end
of the evolution, the difference becomes quite significant. The
disruption time of the $N$-body models is
slightly larger than that of the FP models. This is due to the fact that the escape rate
varies with the number of stars. We perform more detailed
comparisons on the evaporation of stars in section 3.3.
It is easier to determine the
exact core collapse time based on the evolution of the central velocity dispersion
because the fluctuation amplitude is
smaller than that of the central density. The times of core collapse
listed in Table 2 are determined by inspecting the behavior of
$\sigma_c$.


\begin{table}
\begin{minipage}{65mm}
\raggedleft
\caption{Time scales of tidally bound models.}
\vspace{1truemm}
\begin{tabular}{@{}cccccc@{}} \hline\hline
Model         & $W_0$ & $\omega_0$ & $t_{cc}[\tau_{rh,0}]$ &
$t_{dis}[\tau_{rh,0}]$
& $M_{cc}$ \\[3pt] \hline\hline
              &       &     0.0    &  10.1    &  24.30 & 0.63  \\
$N$-body      &  6    &     0.3    &  8.7     &  17.75 & 0.55  \\
              &       &     0.6    &  6.9     &  11.55 & 0.37  \\[1pt] \hline
              &       &     0.0    &  11.73   &  22.61 & 0.59  \\
FOPAX         &  6    &     0.3    &  10.31   &  16.96 & 0.48  \\
              &       &     0.6    &  7.27    &  10.08 & 0.33  \\[1pt] \hline\hline

\end{tabular}
\end{minipage}
\label{tab1}
\end{table}

We show the evolution of the Lagrangian radii of the equal-mass models
in Fig. 4. The results from the FP simulation are displayed by the dashed lines.
Each line represents the radii where the cluster contains $1\%, 5\%,
10\%, 20\%, 50\%$, and $75\%$ of the {\it initial} mass of the cluster.
It is not straightforward to determine the Lagrangian radii in a
flattened system. In Paper \rm{I}, the Lagrangian radii were evaluated
along the specific zenith angle where the effects of
flattening on the mass shells are expected to be less important;
the same zenith angle is used in Fig. 4.
However, as our models are nearly spherical, we determine the Lagrangian
radii on the assumption that the system is spherically symmetric for the
$N$-body models. As seen in Fig. 4, differences between the FP and the
$N$-body models are very small. Analyzing in more depth, we notice
that after the core collapse, the inner part of the cluster expands
more rapidly in the FP model than in the $N$-body model, although the difference is rather
small.

\begin{figure}
\epsfig{figure=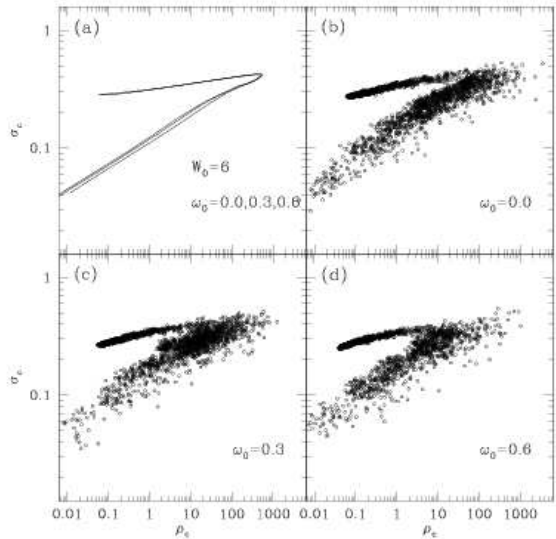, height=0.490\textwidth, width=0.490\textwidth}
\vspace{-3mm}
\caption{The evolution of $\sigma_c$ as a function of $\rho_c$. $\sigma_c$ follows
power laws during pre- and post-collapse states. The evolution is nearly independent of $\omega_0$.}
\label{fig_rhocsigc}
\end{figure}


The relationship between the central density and the central velocity
dispersion is shown in Fig. 5. The upper-left panel
(Fig. 5a) shows the relation between  $\sigma_c$ and $\rho_c$ which is obtained
from all the three FP models with different initial rotations. The other
three panels (Figs. 5b, 5c, and 5d) show the results of the $N$-body models. From
Papers I and II, we know that this relationship is not affected considerably
by the initial rotation, as shown in Fig. 5a. Again, we find
good agreement between the FP and the $N$-body models. The large amount of scatters during the
post core collapse phase, as shown in Fig. 5 is mainly due to the large fluctuation
in the central density. The power-law behavior of $\sigma_c$ on $\rho_c$
during the pre-collapse phase is a consequence of the
self-similarity of the collapsing core and it is well known that
$\sigma_c \propto \rho_{c}^{0.1}$ during this stage (Cohn 1980).
During the post-collapse phase, we can derive the relationship
between $\sigma_c$ and $\rho_{c}$ using an energy balance argument and the
assumption of self-similar evolution (see section 3.4). It follows
that $\sigma_c \propto \rho_{c}^\beta$ with $\beta=0.25$, which is
in good agreement with $\beta=0.23$ derived from the FP model in Paper \rm
{II}. The $N$-body results appear to follow similar power-law behavior, although
the power-law index $\beta$ is difficult to determine because of
the large scatter.

From the $N$-body calculations, we have confirmed the earlier
finding of a significant acceleration in cluster evolution due to
rotation, which was obtained from the FP calculations in Papers \rm {I} and \rm {II}. We also
find that both the $N$-body and FP models give similar results, although there
are small differences with regard to the time of core collapse, and the
disruption times.

\subsection{Evolutions of anisotropy and angular momentum}
\medskip

In axially symmetric systems, the natural decomposition of
velocity vectors is to use the cylindrical coordinate which has
its origin at the center of mass of the cluster.
We investigate the evolution of the velocity dispersions ($\sigma_R$, $\sigma_{\phi}$, and
$\sigma_z$) and
show the evolution of these quantities in Fig. 6,
where $(R,\phi,z)$ represents the conventional axis of the cylindrical
coordinate system.

\begin{figure}
\epsfig{figure=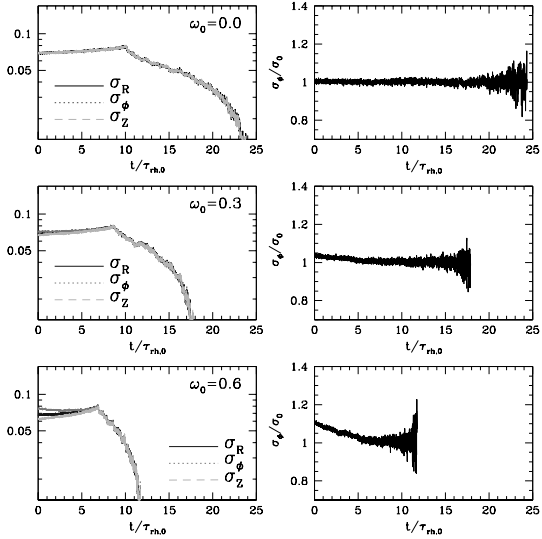, height=0.49\textwidth, width=0.49\textwidth}
\vspace{-3mm}
\caption{Time evolutions of $\sigma_R$, $\sigma_{\phi}$, and $\sigma_Z$(left panels) and
$\sigma_{\phi}/\sigma_0$(right panels) for each models. At the time of core collapse,
$\sigma_R$, $\sigma_{\phi}$, and $\sigma_Z$ are almost equal and
$\sigma_{\phi}/\sigma_0$ is $\approx 1$, where $\sigma_0$ is the 1D velocity dispersion.}
\label{Fig 5.}
\end{figure}

\begin{figure}
\epsfig{figure=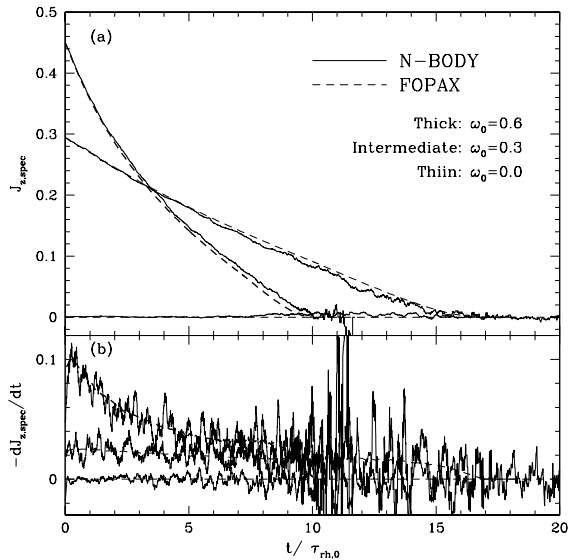, height=0.490\textwidth, width=0.490\textwidth}
\vspace{-3mm}
\caption{
Evolution of the $z$-component of angular momentum per unit mass ($J_z$) (Fig. 7a) and
the loss rate of $J_z$ (Fig. 7b).
The results for the $N$-body model and FP model are shown with solid lines and
dashed lines, respectively. The $J_z$ evolution for the non-rotating model
indicates the corresponding error in determining the $J_z$.
It is clearly shown that $J_z$ for rotating models decreases with time due to the escape of
stars possessing an angular momentum and that the time evolutions of $J_z$ in the
$N$-body and FP models agrees quite well. The loss rate of $J_z$ for the $N$-body models
(solid lines) is smoothed for easier comparison.
}
\label{fig_jzspec}
\end{figure}

For initially rotating models, these three velocity dispersions have
different values. In the right panels of Fig. 6, the ratio of
$\sigma_{\phi}$ to $\sigma_0$
is initially greater than 1 for the rotating models and approaches the isotropic
value of 1, where $\sigma_0$ represents the
average 1D velocity dispersion defined by
$\sigma_0 = (\sigma_R^2+\sigma_\phi^2+\sigma_z^2)^{1/2}/\sqrt{3}$.

The angular momentum in rotating stellar systems is transferred outward through
two-body relaxation, and is also lost due to escaping stars. The stars
gaining a large angular momentum migrate to the outer parts of the system,
while those losing angular momentum drift toward the central parts. As
the stars with a high angular momentum move outward and finally escape
from the cluster, the total angular momentum of the system decreases with
time.

In Fig. 7, we display the evolutions of the $z$-component of the angular
momentum per unit mass ($J_z$) of an entire cluster. The results from the
$N$-body models are shown by solid lines, while those from the FP models are shown by
dashed lines. To indicate the degree of error in determining $J_z$,
we also show the time evolution of $J_z$ for the non-rotating model.
We first note that there is good agreement between the $N$-body and
FP results. The effect of neglecting the third integral in the FP models is
minimized when dealing with global cluster properties (e.g., total
cluster mass and total angular momentum), while this effect is severe for
local properties (e.g., central density).
It is clearly shown that $J_z$ monotonically decreases with
time. The loss rate of angular momentum is large in the early phases and decreases
as the cluster evolves (see Fig. 7b).
The combined effect of gravitation and
rotation accelerates the evolution of the cluster.
A substantial
loss of the initial angular momentum in an entire cluster prevents rotation from
playing an important role in the evolution of a cluster in later phases.
Since the cluster is losing mass at a nearly constant rate,
the total angular momentum of the cluster decreases more rapidly
than $J_z$.

\begin{figure}
\epsfig{figure=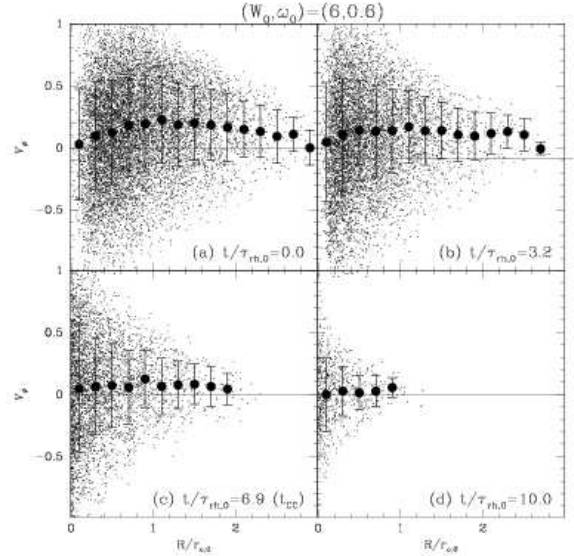, height=0.490\textwidth, width=0.490\textwidth}
\vspace{-3mm}
\caption{
Distribution of the $V_\phi$ at 4 selected
evolutionary epochs in cylindrical coordinate
system for a model with $(W_0,\omega_0)=(6,0.6)$. When there is no rotation,
it should show a symmetry with respect to $V_{\phi} = 0$.
A asymmetry of $V_{\phi}$ disappears with
increasing time due to the loss of the angular momentum. The filled circles are the average
$V_\phi$ values and the error bar corresponds to 1 $\sigma$ dispersion.}
\label{fig_rvphi}
\end{figure}


\begin{figure}
\epsfig{figure=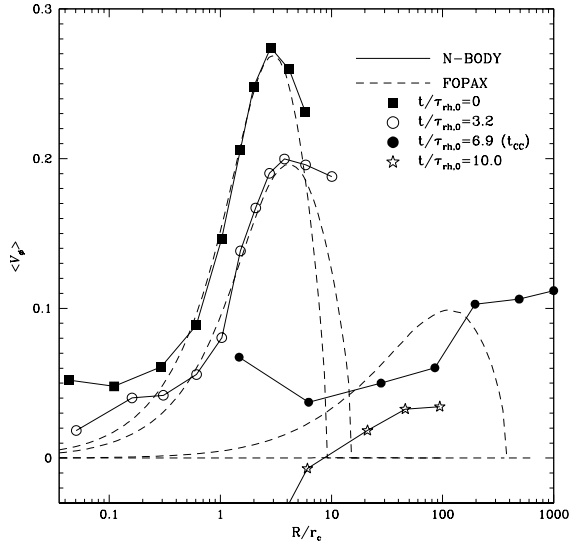, height=0.490\textwidth, width=0.490\textwidth}
\vspace{-3mm}
\caption{
Radial profiles of mean rotational velocities of the cluster model with
$(W_0,\omega_0) = (6,0.6)$ at four evolutionary stages. The radii are
measured in units of current core radius ($r_c$). For comparison, we
show the rotational profiles from the FP models by dashed lines.
}
\label{fig_vrot}
\end{figure}

\begin{figure}
\epsfig{figure=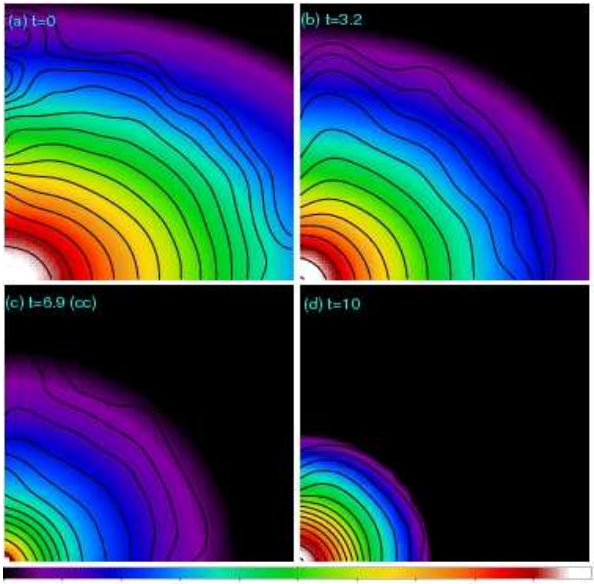, height=0.490\textwidth, width=0.490\textwidth}
\vspace{-3mm}
\caption{
Density distribution of stars in ($R,Z$) coordinate at four selected evolutionary
epochs for a model with $(W_0,\omega_0)=(6,0.6)$ as computed by N-body (contour map) and
FP methods (color map). The horizontal axis is $R$ running from 0 to 3 in
units of the initial core radius, while the vertical axis is the $z$ coordinate
with the same scale. The $N$-body and FP methods give almost the same
density distribution.}
\label{fig_2dcont}
\end{figure}


\subsection{Evolutions of rotation curve}
\medskip
We can construct the rotation curves by computing the averages of $V_\phi$, which is the
$\phi$-component of the velocity of stars. We choose the model
with $\omega_0 = 0.6$ and present the distribution of
$V_\phi$ for all stars at some specific epochs
in Fig. 8. The asymmetry of $V_{\phi}$ with respect to $V_{\phi} =
0$ indicates the global rotation of the stellar
system. During the pre-collapse phase, $V_{\phi}$ becomes more dispersive, particularly
around the central region. A large spread
near the center (R $\sim$ 0) at $t=t_{cc}$ and the slight
asymmetry of $V_{\phi}$ indicate that the cluster is losing
angular momentum; this also indicates that as the core collapse approaches,
the stars are falling into the central region of
the cluster with a high angular speed (but low rotation velocity).
After $t_{cc}$, $V_{\phi}$ still has a small amount of asymmetry.

The radial profiles of the rotational velocity ($v_{rot}$) for the model
with $\omega_0=0.6$ are shown in Fig. 9. We also display the
radial profiles of $v_{rot}$ obtained from the FP model by dashed lines.
The profiles at different evolutionary epochs are shown with
different symbols: filled squares for the initial model, open circles for
the pre-collapse phase, filled circles at the time of core collapse, and
open star marks for the post-collapse phase. The radial profiles
in Fig. 9 are obtained by averaging $V_\phi$ for
stars having the same value of $R$ in cylindrical coordinates.

The radial profiles of $v_{rot}$ at the initial epoch shows
good agreement between the $N$-body and FP models except for a very
central region ($R<0.5 r_c$).
The amount of rotation decreases with
time and the radius where $v_{rot}$ has the peak value progresses outward when
the radius is measured in units of current core radius. This is a consequence of the
self-similar evolution during the pre-collapse phase (Papers I \& II).
The radial profiles of $v_{rot}$ for the $N$-body model during pre-collapse phaes
agree well with those for the FP model. However, at later times, the two methods
yield somewhat different rotation curves. For example, at the time of core collapse,
the $N$-body model predicts a larger
rotation speed for the stars in the central
and outer parts as compared to the FP method.
The rotation seems to continue until a very late epoch ($t=10 t_{rh,0}$)
in the $N$-body model, while there is virtually no rotation left in the FP model.

In Fig. 10, we have shown the distribution of the stellar density in the form of a
color map for the FP results and in the form of isodensity contours for
the $N$-body results. The four panels represent different epochs for the cluster with
largest initial rotation speed ($\omega_0=0.6$).
We display the cluster shapes
in $(R,z)$ coordinates because we assume that the initial model has the
axis of rotation along $z$ in cylindrical coordinates. To compare the cluster shapes
at different epochs, we measure the size in units of initial core radius ($r_{c,0}$).
First, we notice that the shapes computed by the $N$-body and FP models are almost identical.
As the rotation becomes negligible, the stellar system becomes more spherical. The
effects of rotation on the flattened shape of a cluster are observed
during the time of core collapse and the post-collapse phase (Figs. 10c \& 10d);
the shape of the cluster around the central region is almost spherical.

\subsection{Tidal boundary}
\medskip

If the cluster rotates around the host galaxy on a circular orbit, the
tidal field experienced by the cluster does not change with time.
In a steady tidal field, the tidal radius is expressed as follows:
\begin{equation}
r_{tid} \approx  \left( \frac{M}{3M_G} \right)^{1/3} R_{G},
\label{eq-trhi}
\end{equation}
where $M$ is the total mass of the cluster within the tidal boundary;
$R_{G}$, the distance of the cluster from the galactic center; and $M_G$, the
galactic mass within $R_G$. The above equation ensures that the
mean density within the tidal radius is a constant. As the cluster loses
the stars beyond the tidal boundary, the tidal radius has been
adjusted to maintain a constant mean density.

\begin{figure}
\epsfig{figure=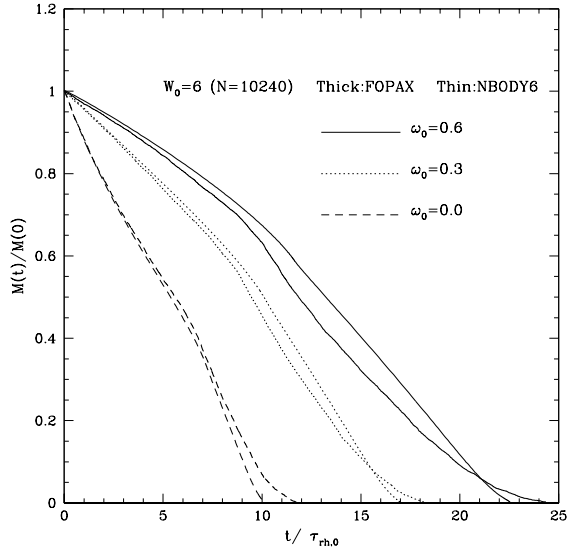, height=0.490\textwidth, width=0.490\textwidth}
\vspace{-3mm}
\caption{The evolution of the total mass of the cluster. The discrepancy between FOPAX and NBODY6
is mainly due to the number of particles after the core collapse.}
\label{Fig 9.}
\end{figure}

We depict the evolution of the total mass of the cluster in Fig. 11.
During the pre-collapse phase, the evolution of the total mass, which is computed
from the $N$-body and FP models, are similar but shows small deviation each other.
However, after the core collapse, the two methods result in somewhat different
bevaiour. This descrepancy in post-collapse phase is partly due to the accumulation
of small difference during the pre-collapse phase.
The escape rate of the $N$-body simulation is lower than that
of the FP simulation. This may be caused due to the decrease in the number of stars in
the cluster. We have assumed the {\it instantaneous} escape of stars
whose energy exceeds the tidal energy. However, in the FP simulations,
the time step is determined by the fraction of the relaxation
time while the $N$-body uses a time step proportional to the {\it
crossing time scale}. As the cluster loses its mass, the ratio between
these two time steps changes. In other words, the crossing time scale
and the relaxation time scale have the following relationship (Binney \&
Tremaine 1987):

\begin{equation}
t_{relax} \approx  \frac{0.1N}{\ln \Lambda} t_{cross}.
\label{eq-trhi}
\end{equation}
Therefore, $t_{cross}/t_{relax}$ increases as
$N$ decreases. This means that the $N$-body simulation removes stars less effectively
than the FP simulation for small value of $N$; this is why we observe 
a long tail in the $N$-body results in
Fig. 11. In $N$-body simulations, after the core collapse, the number of stars remaining
in the cluster is of the order of $10^3$. This is not a sufficient number to
obtain results comparable with the FP method. Hence, the escape rate for
$N$-body models is lesser than that for the FP models during the late stages of the evolution and
$N$-body models survive longer than FP models. We can also observe this
effect in Fig. 3. In this figure, central density features are observed to be
inconsistent between the $N$-body and the FP models
toward the end of the evolution. With more stars, this gap would become
narrower. Since the number of stars in a real globular cluster is considerably
larger than that used in the present $N$-body simulations, the difference
between $N$-body and FP models would decrease for a realistic number of stars
(see Takahashi \& Portegies-Zwart 1998).

\subsection{Core, half-mass, and tidal radii}
\medskip

\begin{figure}
\epsfig{figure=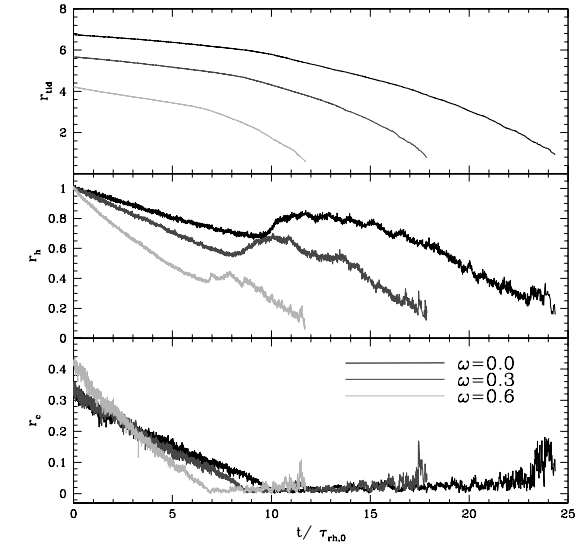, height=0.490\textwidth, width=0.490\textwidth}
\vspace{-3mm}
\caption{Time evolution of $r_c$, $r_h$, and $r_{tid}$ measured in units of
initial half-mass radius ($r_{h,0}$).
The most rapidly rotating cluster has the largest $r_c$ and $r_h$, and the smallest $r_{tid}$.}
\label{fig_rad1}
\end{figure}

\begin{figure}
\epsfig{figure=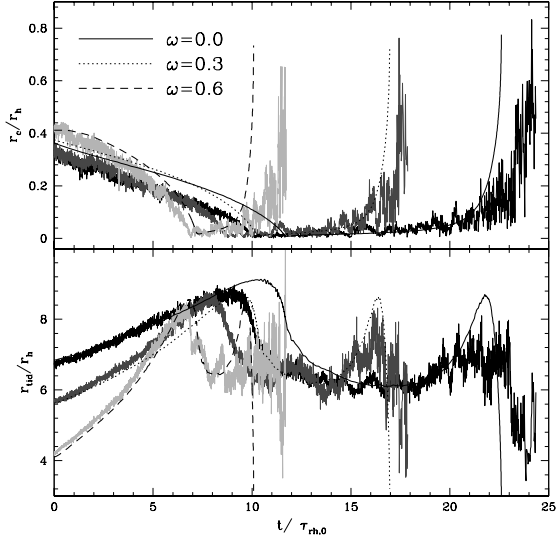, height=0.490\textwidth, width=0.490\textwidth}
\vspace{-3mm}
\caption{Time evolution of $r_c/r_h$ and $r_{tid}/r_h$. Self-similarity feature after $t_{cc}$ is shown.
In particular, the evolution of $r_{tid}/r_h$ shows good agreement with that of FP.}
\label{fig_rad2}
\end{figure}

We now investigate the evolution of the core, half-mass, and tidal
radii of the star cluster. The behavior of $r_{tid}$ and $r_c$ in
units of initial half-mass radius ($r_{h,0}$) is shown in Fig.
12. As the rapidly rotating initial model has a 
half-mass radius smaller than the slowly rotating model, the initial value of
$r_c/r_h$ for the most rapidly rotating cluster is larger than that of the
other two models, as shown in Fig. 13 (Paper II).
The evolutions of $r_c$ and $r_{tid}$ in units of $r_h$
at the same evolutionary stage as those in Fig. 12
are shown in Fig. 13. We find that there are significant differences
in $r_c/r_h$ between the FP and the $N$-body models, while the difference is not so
apparent in $r_{tid}/r_h$. This reflects the difficulties in
determining $r_c$ for the $N$-body models with a relatively small value of $N$ rather
than any systematic differences in different approaches. Both $r_c$
and $r_h$ decrease with time, although there is a difference in the
decreasing rate that depends on the initial degree of rotation.
After the core collapse, both $r_c$ and $r_h$ increase for some time at
almost the same rate due to self-similar expansion. Subsequently, the tidal
boundary shrinks rapidly after core bounce. Therefore, the half-mass
radius begins to decrease again. However, the 
shrinking of $r_{tid}$ does not affect the central region and $r_c$
continues to increase. At the end of the evolution $r_c$ shows a steep
increase; this signals the complete disruption of the cluster.

After the core collapse, $r_c/r_h$ and $r_{tid}/r_h$ show nearly the same
behavior. The value of $r_c/r_h$ is almost constant and the
evolution of $r_{tid}/r_h$ for different initial models is similar to each other and
independent of the rotation parameter $\omega_0$. We already have
assumed the self-similar evolution of the inner part of the
stellar system in order to explain the relation between $\rho_c$
and $\sigma_c$ after $t_{cc}$ (Fig. 4). We can express $\sigma_c^2
\sim \frac{M_c}{r_c}$ and $\sigma_h^2 \sim \frac{M_h}{r_h}$, where
$M_c$ and $M_h$ are the masses within $r_c$ and $r_h$, respectively.
Since the inner parts of the cluster are nearly isothermal,
we obtain $\sigma_h/\sigma_c = constant$. With the self-similarity
assumption ($r_c/r_h=constant$), we can rewrite $\sigma_c$ as
$\sigma_c \sim \rho_h^{1/6} M^{1/3}$.
On the other hand, according to Goodman (1987), the energy balance argument
predicts that $\rho_c/\rho_h \propto M^{4/3}$.
Therefore, $\sigma_c \sim \rho_c^{1/4} \rho_h^{-1/12}$.
If we use the tidal boundary condition, $M/r_{tid}^{3} = constant$ and
$\rho_h \sim \frac{M}{r_h^3}$, we can obtain the following relation:
\begin{equation}
\ln \sigma_c \sim \frac{1}{4} \ln \rho_c - \frac{1}{4} \ln \frac{r_{tid}}{r_h} .
\label{eq-sigrho}
\end{equation}
The variation in $\frac{r_{tid}}{r_h}$ during the post-collapse phase is
very small (by a factor of few) as compared with that in $\rho_c$ (by
a few orders of magnitude), except  near the disruption time.
Therefore, we can approximately write following relation:
\begin{equation}
\ln \sigma_c \sim \frac{1}{4} \ln \rho_c.
\label{eq-sigrho2}
\end{equation}
If we express $\sigma_c \propto \rho_c^{\beta}$, $\beta=0.25$, then.
This value is close to $0.23$ that was achieved in Paper \rm {II}.

\section{Summary and discussion}
\medskip

We have performed numerical simulations for the evolution of
initially rotating star clusters with equal mass using NBODY6 and have compared
the results with those computed by the direct integration of the Fokker-Planck
equation. We have considered clusters with $N=10240$.
For critical comparisons between $N$-body and FP models, we
constructed the initial $N$-body models using the initial 2D distribution
function used for FP models.

We observed the acceleration of the core collapse,
as reported in Papers \rm{I} \& \rm {II}. The degree of
acceleration obtained from the present $N$-body models is slightly different from
that obtained from the FP models; however, the small difference in the core-collapse
time between the $N$-body and statistical methods (FP model, gaseous
model, etc.) has also been observed earlier (Spurzem \& Aarseth 1996). The
entire evolutionary trend of the central density agrees with that of the
FP models.


The $z$-component of the specific angular momentum ($J_z$)
is observed to monotonically decrease
with time for the clusters with initial rotation. The global evolutionary
trend of $J_z$ between the $N$-body and the FP models shows excellent
agreement.
The loss rate of $J_z$ decreases as the cluster evolves.
Therefore, we conclude that during the early stages the existence of
initial rotation significantly affects
the entire cluster evolution.


In FP simulations, the cluster evolution will be
independent of the third integral to the end of time. On the other
hand, in the $N$-body simulations, the third integral effect may appear
during the evolution. In addition, there is a limit on the number of stars
and the random fluctuations of the $N$-body models in this study and these
limits also causes differences with the FP method. Therefore, we need to
perform more $N$-body simulations with a larger number of stars or
with various models by using different random number.

\section*{Acknowledgments}

The authors would like to thank the anonymous referee for his/her helpful comment.
HML is supported by KRF grant No. 2006-341-C00018.

\end{document}